%% file: Lattice2018proc.tex
\documentclass{PoS}%
\usepackage{amsmath,amssymb}
\usepackage{bm}
\usepackage{mathrsfs}
\usepackage{cleveref}
\usepackage{braket}
\usepackage{amsmath}
\usepackage{amsfonts}
\usepackage{amssymb}
\usepackage{graphicx}
\usepackage{color}%
\providecommand{\U}[1]{\protect\rule{.1in}{.1in}}

\input{front-matter.txt}

\begin{document}
\section{Introduction}

The dual superconductor picture is one of the most promising scenarios for
quark confinement \cite{dualsuper}. According to this picture, magnetic
monopoles causing the dual superconductivity are regarded as the dominant
degrees of freedom responsible for confinement. However, it is not so easy to
establish this hypothesis. Indeed, even the definition of magnetic monopoles
in the pure Yang-Mills theory is not obvious. If magnetic charges are naively
defined from electric ones by exchanging the role of the magnetic field and
electric one according to the electric-magnetic duality, one needs to
introduce singularities to obtain non-vanishing magnetic charges, as
represented by the Dirac monopole. For such configuration, however, the energy
becomes divergent.

There are two prescriptions avoiding this issue in defining magnetic
monopoles, i.e., \emph{the Abelian projection method} and \emph{ the
decomposition method,} In the\emph{ }Abelian projection method \cite{Hooft81},
the \textquotedblleft diagonal component\textquotedblright\ of the Yang-Mills
gauge field is identified with the Abelian gauge field, and a magnetic
monopole is defined as the Dirac monopole. The energy density of this monopole
can be finite everywhere because the contribution from the singularity of a
Dirac monopole can be canceled by that of the off-diagonal components of the
gauge field. However, one needs to fix the gauge, because otherwise the
\textquotedblleft diagonal component\textquotedblright\ is meaningless. In the
decomposition method, on the other hand, we need no more gauge fixing. The
gauge-covarint decomposition was first proposed by Cho, Duan and Ge \cite{CDG}
for the $SU(2)$ Yang-Mills theory, and was extended to the $SU(N)$ Yang-Mills
theory (see \cite{KKSS15} for review). The key ingredient in this
decomposition is the Lie-algebra valued field with a unit length which we call
the \textit{color field}. Then, the monopoles can be defined by using the
gauge-invariant part proportional to the color field in the field strength
just like the Abelian field strength in the Abelian projection. Therefore, the
gauge invariance is explicitly maintained in the decomposition method.

It should be examined on the lattice whether or not these monopoles can
reproduce the expected infrared behavior of the original Wilson loop average.
In the preceding lattice studies for $SU(2)$ and $SU(3)$ Yang-Mills theory
within the Abelian projection method using the maximal Abelian (MA) gauge, it
was confirmed that (i) the string tension calculated from the diagonal part of
the original Yang-Mills field reproduces the full string tension calculated
from the original Yang-Mills fields \cite{SY90,STW02}, and that (ii) the
string tension calculated from the magnetic monopole extracted from the
diagonal part mostly reproduce the full string tension \cite{SS94,STW02}.
However, it is known that the resulting monopole contribution does not
reproduce the original Wilson loop average if the Abelian projection is
naively applied to the Wilson loop in higher representations \cite{DFGO96}.
This is because, in higher representations, the diagonal part of the Wilson
loop does not behave in the same way as the original Wilson loop. Poulis
heuristically found the correct way to extend the Abelian projection approach
for the adjoint representation in the $SU(2)$ Yang-Mills theory
\cite{Poulis96}.

In this talk, we propose a systematic prescription to extract the
\textquotedblleft dominant\textquotedblright\ part of the Wilson loop average,
which can be applied to the Wilson loop operator in an arbitrary
representation of the $SU(N)$ Yang-Mills theory. To test this proposal, we
perform the numerical simulation on the lattice, and measure the Wilson loop
average in the higher representation for both the original Yang-Mills field
and the restricted field which is extracted as the dominant mode for quark
confinement by using the decomposition method, e.g., the Wilson loop averages
for the adjoint representation in the $SU(2)$ Yang-Mills theory, and for the
adjoint and sextet representations in the $SU(3)$Yang-Mills theory. We find
that the results support our claim.

\section{The gauge field decomposition method and the non-Abelian Stokes
theorem}

In this section, we give a brief review of the gauge field decomposition (
e.g., see \cite{KKSS15}) and the non-Abelian Stokes theorem for the Wilson
loop in an arbitrary representation \cite{MK15}.

\subsection{The gauge filed decomposition}

We decompose the gauge link variable $U_{x,\mu}$ into the product of the two
variables $V_{x,\mu}$ and $X_{x,\mu}$ in such a way that the new variable
$V_{x,\mu}$, is transformed by the full $SU(N)$ gauge transformation
$\Omega_{x}$ as the gauge link variable $U_{x,\mu}$, while $X_{x,\mu}$
transforms as the site variable:
\begin{subequations}
\begin{align}
&  X_{x,\mu},U_{x,\mu}=X_{x,\mu}V_{x,\mu}\in G=SU(N),\\
&  U_{x,\mu}\longrightarrow U_{x,\nu}^{\prime}=\Omega_{x}U_{x,\mu}%
\Omega_{x+\mu}^{\dag},\text{ \ \ }V_{x,\mu}\longrightarrow V_{x,\nu}^{\prime
}=\Omega_{x}V_{x,\mu}\Omega_{x+\mu}^{\dag},\text{ \ }X_{x,\mu}\longrightarrow
X_{x,\nu}^{\prime}=\Omega_{x}X_{x,\mu}\Omega_{x}^{\dag}.
\end{align}
From the physical point of view, $V_{x,\mu}$, which we call the restriced
field, could be the dominant mode for quark confinement, while $X_{x,\mu}$ is
the remainder part. The possible options of the decomposition are
discriminated by the stability subgroup of the gauge group. Here, we only
consider the maximal option.

The maximal option is obtained for the stability subgroup of the maximal torus
subgroup of $G$: $\tilde{H}=U(1)^{N-1}\subset SU(N).$ The resulting
decomposition is the gauge-invariant extension of the Abelian projection in
the maximal Abelian (MA) gauge. \ We introduce color fields as,
\end{subequations}
\begin{equation}%
\bm{n}%
^{(k)}(x)=\Theta(x)H_{k}\Theta^{\dag}(x)\in Lie[G/\tilde{H}]\text{
\ \ \ \ \ \ }(k=1,\ldots,N-1)\text{,}%
\end{equation}
which are expressed using a common $SU(N)$-valued field $\Theta(x)$ with the
Cartan generators $H_{k}$. The decomposition is obtained by solving the
defining equations:
\begin{subequations}
\begin{align}
D_{\mu}^{\epsilon}[V]\mathbf{n}_{x}^{(k)} &  :=\frac{1}{\epsilon}\left[
V_{x,\mu}\mathbf{n}_{x+\mu}^{(k)}-\mathbf{n}_{x}^{(k)}V_{x,\mu}\right]
=0\text{ },\label{eq:define-max}\\
\text{\ }g_{x} &  :=e^{i2\pi q/N}\exp\left(  -i\sum_{j=1}^{N-1}a_{x}%
^{(j)}\mathbf{n}_{x}^{(j)}\right)  \text{,}\label{eq:define-resdidual}%
\end{align}
where, the variable $g_{x}$ is the $U(1)^{N}$ part which is undetermined from
Eq.(\ref{eq:define-max}) alone, $a_{x}^{(j)}$ are coeficients, and $q$ is an
integer. Note that the above defining equations correspond to the continuum
version ($g_{x}=\mathbf{1}$): $D_{\mu}[%
\mathscr
V]\mathbf{n}^{(k)}(x)=0$ and $\mathrm{tr}(%
\mathscr
X_{\mu}(x)\mathbf{n}^{(k)}(x))$ $=0$, respectively. These defining equations
can be solved exactly, and the solution is given by
\end{subequations}
\begin{align}
X_{x,\mu} &  =\widehat{K}_{x,\mu}^{\dag}\det(\widehat{K}_{x,\mu})^{1/N}%
g_{x}^{-1},\text{ \ \ \ \ \ \ \ \ \ }V_{x,\mu}=X_{x,\mu}^{\dag}U_{x,\mu
},\nonumber\\
\widehat{K}_{x,\mu} &  :=\left(  K_{x,\mu}K_{x,\mu}^{\dag}\right)
^{-1/2}K_{x,\mu},\text{ \ \ \ \ \ \ \ \ \ \ \ }K_{x,\mu}:=\bm1+2N\sum
_{k=1}^{N-1}\bm n_{x}^{(k)}U_{x,\mu}\mathbf{n}_{x+\mu}^{(k)}U_{x,\mu}^{\dag
}.\label{eq:decomp-max}%
\end{align}
In the naive continuum limit, we can reproduce the decomposition in the
continuum theory.

The color fields $%
\bm{n}%
^{(k)}$ are obtained by minimizing the functional $R_{\mathrm{MA}}$ for a
given configuration of the link variables $U_{x,\mu}$ with respect to the
gauge transformation
\begin{equation}
R_{\mathrm{MA}}[U,\{\bm n^{(k)}\}]=\sum_{x,\mu}\sum_{k=1}^{N-1}%
\operatorname{tr}[(D_{\mu}[U]\mathbf{n}_{x}^{(k)})^{\dag}(D_{\mu}%
[U]\mathbf{n}_{x}^{(k)})]. \label{red_con}%
\end{equation}
If we choose this condition as the reduction condition, the definition of
monopoles is equivalent to that for the Abelian projection in the MA gauge. In
the present study for the $SU(3)$ gauge theory, we apply two additional
reduction conditions which are defined by minimizing following functionals
\begin{align}
R_{n3}[U,\{\bm n^{(k)}\}]  &  =\sum_{x,\mu}\operatorname{tr}[(D_{\mu
}[U]\bm n_{x}^{3})^{\dag}(D_{\mu}[U]\bm n_{x}^{3})],\quad(\bm n_{x}%
^{3}:=\Theta_{x}T^{3}\Theta_{x}^{\dag}),\label{n3}\\
R_{n8}[U,\{\bm n^{(k)}\}]  &  =\sum_{x,\mu}\operatorname{tr}[(D_{\mu
}[U]\bm n_{x}^{8})^{\dag}D_{\mu}[U]\bm n_{x}^{8})],\quad(\bm n_{x}^{8}%
:=\Theta_{x}T^{8}\Theta_{x}^{\dag}). \label{n8}%
\end{align}

\subsection{Non-Abelian Stokes theorem}

We can relate the decomposed field variables to a Wilson loop operator through
a version of the non-Abelian Stokes theorem (NAST) which was proposed by
Diakonov and Petrov \cite{DP89}. In this version of the NAST, a Wilson loop
operator in a representation $R$ is rewritten into the surface integral form
by introducing a functional integral on the surface $S$ surrounded by the loop
$C$ as
\begin{subequations}
\begin{align}
W_{R}[\mathscr A;C]  &  :=\int D\Omega\exp\left(  ig\int_{S:\partial
S=C}dS_{\mu\nu}\sum_{k=1}^{N-1}\Lambda_{k}F_{\mu\nu}^{(k)}\right)  \text{
},\label{NAST}\\
F_{\mu\nu}^{(k)}  &  :=2\operatorname{tr}(\bm n^{(k)}\mathscr F_{\mu\nu
}[\mathscr V]),
\end{align}
where $D\Omega$ is the product of the Haar measure over the surface $S$,
$\Lambda_{k}$ is the $k$-th component of the highest-weight of the
representation $R$, the color fields are defined by $\bm n^{(k)}=\Omega
H_{k}\Omega^{\dag}$, and $\mathscr F_{\mu\nu}[\mathscr V]$ is the field
strength for the restricted field $\mathscr V$ \ in the continuous version.
Monopoles are defined in the same manner as the Dirac monopoles for the
Abelian-like gauge-invariant field strength $F_{\mu\nu}^{(k)}$. The resulting
monopoles are gauge invariant by construction. Thus we can relate the
restricted field to tied Wilson loop operator in the manifestly
gauge-invariant way.

In the actual lattice simulations, we do not follow this NAST directly.
Without performing the integration over the measure $D\Omega$, the argument of
the exponential is approximated by substituting the color field (as a
functional of the gauge field) which is obtained by solving the reduction
condition. Indeed, this approximation is commonly used for the Wilson loop in
the fundamental representation, e.g., the Abelian projection in the MA gauge
is equivalent to the field decomposition with the color fields determined by
minimizing the functional \cref{red_con}.

The NAST can be applied not only to the fundamental representation but also to
any representation, which suggests the correct way to extract the dominant
part of the Wilson loop in higher representations as we explain in the next section.

\section{Wilson loops in higher representations}

In the preceding study, it was shown that the area law of the average of the
Wilson loop in the fundamental representation is reproduced by the monopole
contribution. However there is a possibility that the monopole contribution
accidentally coincides with the behavior of the original Wilson loop, and
therefore we should examine the other quantities. The Wilson loops in higher
representations are appropriate for this purpose because they have clear
physical meaning and show the characteristic behavior, e.g., the Casimir scaling.

It is known that if we adopt the Abelian projection naively to higher
representations, the monopole contributions do not reproduce the correct
behavior \cite{DFGO96}. Thus, we have to find a more appropriate way to
extract the monopole contributions. The NAST (\ref{NAST}) suggests the
different operator as the dominant part of the Wilson loop in higher
representations. In the actual calculation, we approximate the NAST by
calculating the integrand of the NAST using the color fields satisfying the
reduction condition. In this approximation, the operator to be calculated is
not equivalent to the Wilson loop for the restricted field. From this point of
view, therefore, the distinct operator is suggested as the dominant part of
the Wilson loop. We can see the difference between the operator suggested by
the NAST and the Wilson loop for the restricted field by expressing it as the
surface integral form as
\end{subequations}
\begin{equation}
W_{R}[\mathscr V;C]=\frac{1}{D_{R}}\sum_{\mathbf{\mu\in}\Delta_{R}%
}d_{\mathbf{\mu}}\exp\left(  ig\sum_{k=1}^{N-1}\mu_{k}\int_{S}dS^{\alpha\beta
}F_{\alpha\beta}^{(k)}\right)  ,\label{W_rest}%
\end{equation}
where $D_{R}$ is the dimension of the representarion $R,$ $\Delta_{R}$ is set
of the weights of $R$, $\ \mu_{k}$ is the $k$-th component of the weight
$\mathbf{\mu,}$ and $d_{\mu}$ is the multiplicity of$\ \mathbf{\mu},$ which
satisfies $D_{R}=\sum_{\mathbf{\mu\in}\Delta_{R}}d_{\mathbf{\mu}}.$ In the
fundamental representation, all weights are equivalent to the highest weight
under the action of \ the Weyl group, but in higher representations there are
weights that are not equivalent to the highest weight. Therefore this
expression is not equivalent to the integrand of \cref{NAST} in higher
representations. For this reason we should modify \cref{W_rest} so as to
include only the weights that is equivalent to the highest weights.

By using the untraced Wilson loop $V_{C}:=\prod_{\braket{x,\mu}\in C}V_{x,\mu
}$ for the restricted field in the fundamental representation, we obtain the
Wilson loop in higher representations. For $SU(2)$, we propose the operator
for the spin-$j$ representation as
\begin{equation}
W_{[j]}^{SU(2)}[\mathscr V;C]=\frac{1}{2}\operatorname{tr}(V_{C}%
^{2j}).\label{su2}%
\end{equation}
For $SU(3)$, we propose the operator for the representation with the Dynkin
index $[m_{1},m_{2}]$ as
\begin{equation}
W_{[m_{1},m_{2}]}^{SU(3)}[\mathscr V;C]=%
\begin{cases}
\frac{1}{6}\left(  \operatorname{tr}(V_{C}^{m_{1}})\operatorname{tr}%
(V_{C}^{\dag m_{2}})-\operatorname{tr}(V_{C}^{m_{1}}V_{C}^{\dag m_{2}%
})\right)   & \text{for}\quad m_{1},\text{ }m_{2}>0\\
\frac{1}{3}\operatorname{tr}(V_{C}^{m_{1}}) & \text{for}\quad m_{2}=0\\
\frac{1}{3}\operatorname{tr}(V_{C}^{\dag m_{2}}) & \text{for}\quad m_{1}=0
\end{cases}
.\label{su3}%
\end{equation}

\section{Numerical result}

In order to support our claim that the dominant part of the Wilson loops in
higher representation is the operator suggested by the NAST, by using the
numerical simulation on the lattice, we check whether the string tension from
the Wilson loop in the representation $R$ of the restricted field $V$\ can
reproduce that of the Yang-Mills field or not.

For this purpose, We set up the gauge configuration for the standard Wilson
action at $\beta=2.5$ on the $24^{4}$ lattice for $SU(2)$ case and at
$\beta=6.2$ on the $24^{4}$ lattice for $SU(3)$ case . For $SU(2)$ case, we
prepare $500$ configurations every $100$ sweeps after $3000$ thermalization by
using the heat bath method. For $SU(3)$ case, we prepare $1500$ configurations
every $50$ sweeps after $1000$ thermalization by using pseudo heat bath method
with over-relaxation algorithm ($20$ steps per sweep). In the measurement of
the Wilson loop average we apply the APE smearing technique for $SU(3)$ case
and the hyper-blocking for $SU(2)$ case to reduce noises and reduced the
exciting modes. The number of the smearing steps is determined so that the
ground state overlap is enhanced \cite{BSS95}. We have calculated the Wilson
loop average $W(R,T)$ for the rectangular loop with length $T$ and width $R$
to derive the potential $V(R,T)$ through the formula $V(R,T):=-\log
(W(R,T+1)/W(R,T).$

\begin{figure}[t]
\centering\includegraphics[width=0.30\hsize]{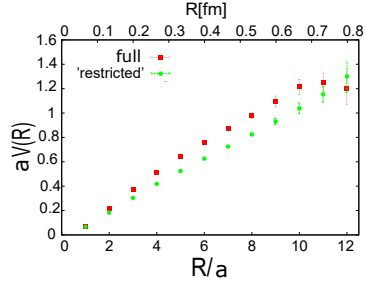}
\caption{ The static potential between the sources in the adjoint
representation of $SU(2)$ using \cref{su2} for $j=1$ and for comparison the
full Wilson loop average in the adjoint representation. The result is
consistent with that of \cite{Poulis96,CHS04} where the same quantity is
calculated by the Abelian projection method.}%
\label{fig:su2}%
\end{figure}In case of the $SU(2)$, we investigate the Wilson loop in the
adjoint representation $\mathbf{3}$ ($j=1$). The restricted field, the
extracted dominant mode for confinement, is obtained by using the
decomposition eq(\ref{eq:decomp-max}) for the color field which minimizes the
reduction condition Eq(\ref{red_con})($N=2).$ Figure \ref{fig:su2} shows that
the string tension from the restricted and Yang-Mills filed in the adjoint
represent are in good agreement. \begin{figure}[t]
\includegraphics[height=45mm, clip, viewport=0 0 198 170 ]{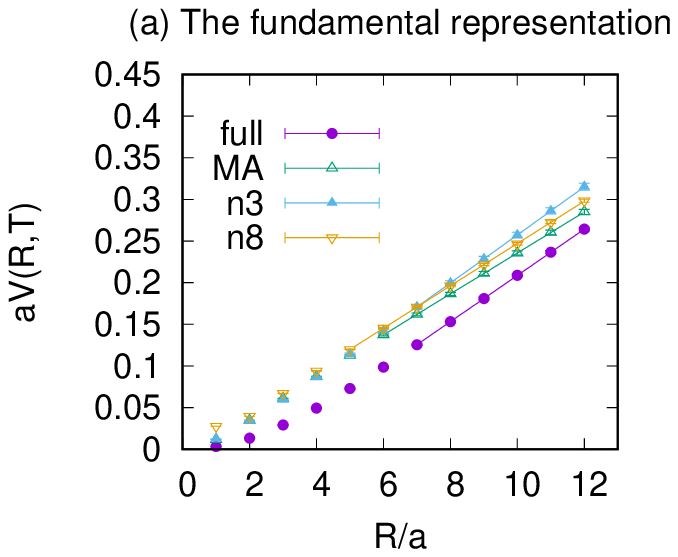}
\includegraphics[height=45mm, clip, viewport= 15 0 198 170 ]{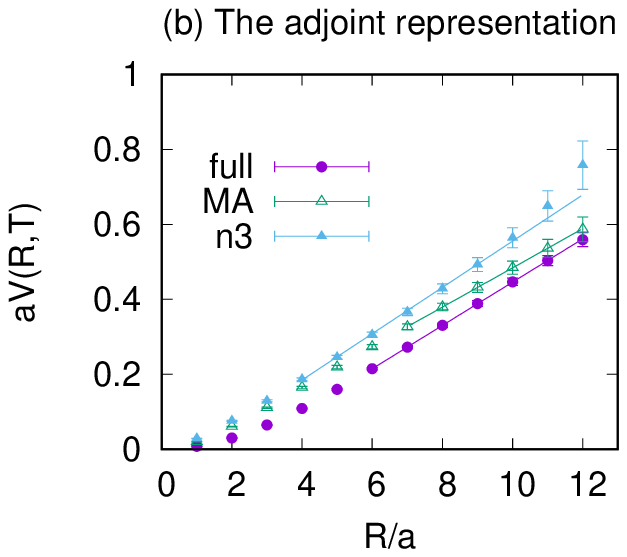}
\includegraphics[height=45mm, clip, viewport=15 0 198 170 ]{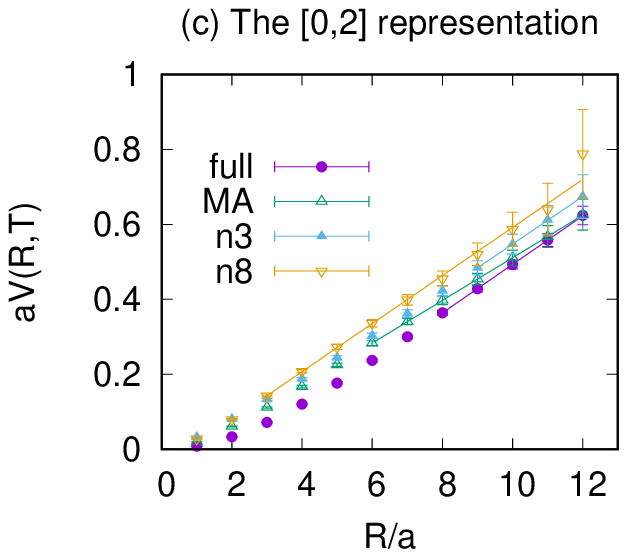}\caption{
The static potential from the Wilson loop average $\left\langle W_{R}%
(R,T=8)\right\rangle $ in the representation $R$ of $SU(3)$ calculated using
\cref{su3}: (a) fundamental $[1,0]$ Rep., (b) adjoint $[1,1]$ Rep. , and
sextet $[0,2]$ Rep. in comparison with the full Wilson loop average. The
legends, MA, n3, and n8, represents the mesurements by using the corresponding
reduction conditions, Eqs(\ref{red_con}), (\ref{n3}), and (\ref{n8}),
respsctively. }%
\label{fig:su3}%
\end{figure}%
\begin{table}[bt] \centering
\begin{tabular}
[c]{c||c|c|c|c}\hline
& full & MA & n3 & n8\\\hline\hline
fundamental & $0.02776(2)$ & $0.02458(1)$ & $0.02884(3)$ & $0.02544(3)$%
\\\hline
adjoint & $0.0576(1)$ & $0.0522(1)$ & $0.062(1)$ & $-$\\\hline
\lbrack0,2] & $0.0647(1)$ & $0.05691(9)$ & $0.0635(2)$ & $0.0641(6)$\\\hline
\end{tabular}
\caption{ The string tensions in the lattice unit in the $SU(3)$ case:
the string tensions obtained under the MA gauge (\ref{red_con}),
and reduction conditions n3 (\ref{n3}) and n8 (\ref{n8}), in comparison with the full string tension. }\label{tbl:sigma-su3}%
\end{table}%

In case of $SU(3)$ we investigate the Wilson loop eq(\ref{su3}), i.e., (a) in
the fundamental representation $[0,1]=\mathbf{3}$, (b) in the adjoint
representation $[1,1]=\mathbf{8}$, and (c)\ in the sextet representation
$[0,2]=\mathbf{6}$. For each representation, we measure the Wilson loop
average for possible reduction conditions, eq(\ref{red_con}) ($N=3)$,
eq(\ref{n3}), and eq(\ref{n8}). \ Figure \ref{fig:su3} shows the static
potentials from the Wilson loop in the higher representations. Table
\ref{tbl:sigma-su3} shows the string tensions which are extracted by fitting
the data with the linear potential for the infrared region. The string
tensions extracted from our proposed "dominant" operators reproduce nearly
equal to or more than $80\%$ of the full string tension. These results
indicate that the proposed operators give actually the dominant part of the
Wilson loop average.

\section{Conclusion}

We have proposed a solution for the problem that the correct behavior of the
Wilson loop in higher representations cannot be reproduced if the restricted
part of the Wilson loop is naively extracted by adapting the Abelian
projection or the field decomposition in the same way as in the fundamental
representation. We have proposed the prescription to construct the operator
suitable for this purpose. We have performed numerical simulations to show
that this prescription works well in the adjoint rep. for $SU(2)$ color group,
and in the fundamental, adjoint , and sextet representations\ for $SU(3)$
color group. Further studies are needed in order to establish the magnetic
monopole dominance in the Wilson loop average for higher representations,
supplementary to the fundamental representation for which the magnetic
monopole dominance was established.

\section*{Acknowledgement}

This work was supported by Grant-in-Aid for Scientific Research, JSPS KAKENHI
Grant Number (C) No.15K05042. R. M. was supported by Grant-in-Aid for JSPS
Research Fellow Grant Number 17J04780. The numerical calculations were in part
supported by the Large Scale Simulation Program No.16/17-20(2016-2017) of High
Energy Accelerator Research Organization (KEK), and were performed in part
using COMA(PACS-IX) at the CCS, University of Tsukuba.

\end{document}

%% file: front-matter.txt
\title{ \begin{flushright}
\begin{minipage}{0.3\textwidth} \normalsize
KEK Preprint 2018-80 \\  CHIBA-EP-234 \\
\end{minipage} \end{flushright} %
Confinement of quarks in higher representations in view of dual superconductivity%
}

\ShortTitle{Confinement of quarks in higher representations in view of dual superconductivity}

\author{\speaker{Akihiro Shibata}\\
Computing Research Center, High Energy Accelerator Research Organization (KEK) \\
SOKENDAI (The Graduate University for Advanced Studies), Tsukuba 305-0801, Japan \\
        E-mail: \email{akihiro.shibata@kek.jp}}

\author{Ryutaro Matsudo\\
Department of Physics, Faculty of Science and Engineering, 
Chiba University, Chiba 263-8522, Japan\\
        E-mail: \email{afca3071@chiba-u.jp}}

\author{Seikou Kato\\
Oyama National College of Technology, Oyama, Tochigi 323-0806, Japan \\
        E-mail: \email{skato@oyama-ct.ac.jp}}

\author{Kei-Ichi Kondo\\
Department of Physics, Graduate School of Science and Engineering, Chiba University, \\
Department of Physics, Graduate School of Science, Chiba University, Chiba 263-8522, Japan \\
        E-mail: \email{kondok@faculty.chiba-u.jp}}

\abstract{
Dual superconductor picture is one of the most promising scenarios for quark confinement. 
We have proposed a new formulation of Yang-Mills theory on the lattice so that the so-called restricted field obtained from the gauge-covariant decomposition plays the dominant role in quark confinement. 
This framework improves the Abelian projection in the gauge-independent manner.
For quarks in the fundamental representation, we have demonstrated  some numerical evidences for the dual superconductivity. 
However, it is known that the expected behavior of the Wilson loop in higher representations cannot be reproduced if the restricted part of the Wilson loop is extracted by adopting the Abelian projection or the field decomposition naively in the same way as in the fundamental representation.
In this talk, therefore, we focus on confinement of quarks in higher representations. 
By virtue of the non-Abelian Stokes theorem for the Wilson loop operator, we propose suitable operators constructed from the restricted field only in the fundamental representation to reproduce the correct behavior of the original Wilson loop in higher representations. 
Moreover, we perform lattice simulations to measure the static potential for quarks in higher representations using the proposed    operators. 
We find that the proposed operators well reproduce the expected behavior of the original Wilson loop average, which overcomes the problem that occurs in naively applying Abelian-projection to the Wilson loop operator for higher representations.
}

\FullConference{The 36th Annual International Symposium on Lattice Field Theory - LATTICE2018\\
		22-28 July, 2018\\
		Michigan State University, East Lansing, Michigan, USA.}